\documentclass{optica-article}

\journal{opticajournal} 

\articletype{Research Article}

\usepackage{braket}
\usepackage{lineno}

\begin{document}

\title{Optimal structured light waves generation in 3D volumes using communication mode optics}

\author{Vinicius S. de Angelis,\authormark{1,2,*} Ahmed H. Dorrah,\authormark{1} Leonardo A. Ambrosio, \authormark{2} David A. B. Miller,\authormark{3,*} and Federico Capasso$^{1}$}

\address{\authormark{1} Harvard John A. Paulson School of Engineering and Applied Sciences, Harvard University, Cambridge, MA 02138, USA \\
\authormark{2}Department of Electrical and Computer Engineering, São Carlos School of Engineering, University of São Paulo, 400 Trabalhador são-carlense Ave., 13566-590, São Carlos, São Paulo, Brazil \\
\authormark{3}Ginzton Laboratory, Stanford University, Stanford, CA 94305, USA \\}

\email{\authormark{*}dabm@stanford.edu / viniciusangelis@seas.harvard.edu} 


\begin{abstract*} 
Achieving precise control of light intensity in 3D volumes is highly in demand in many applications in optics. Various wavefront shaping techniques have been utilized to reconstruct a target amplitude profile within a 3D space. However, these techniques are intrinsically limited by cross-talk and often rely on optimization methods to improve the quality of the reconstruction. We propose and experimentally demonstrate a new wavefront shaping method based on interfering the optimum orthogonal communication modes connecting a source plane and a receiving volume. These optimum modes are computed from the singular value decomposition of a coupling operator that connects each point at the source plane to another one in the receiving volume. The modes comprise a pair of source and receiving eigenfunctions, with each forming a complete orthogonal basis for their respective spaces. We utilize these modes to construct arbitrarily chosen 2D and 3D structured light waves within the output receiving volume and optically generate these waves using a spatial light modulator. Our generated intensity profiles exhibit low cross-talk, high fidelity, and high contrast. We envision our work to inspire new directions in any domain that requires controlling light intensity in 3D with high precision and also to serve as a benchmark for other wavefront shaping techniques.

\end{abstract*}

\section{Introduction}
 
Harnessing the interaction between light and matter shapes our understanding of the universe. Imaging through disorder with high spatial and temporal resolution, for instance, can teach us something new about cells and other galaxies alike. Likewise, light manipulation, storage, and detection are the foundation of photonic-based technologies that include quantum computing, communications, and sensing \cite{2021_wang,He2022,Laucht_2021,Wan2023,Grier2003,KUMAR2021,Lee2023}. At the heart of these developments lies the need for tailoring the properties of incoming light in a nontrivial manner \cite{Genevet_2015,Quevedo-Teruel_2019,HuangZhang,Wan2022,Sedeh2023,Kuznetsov2024}. Early pursuits in structured light waves include the development of custom two-dimensional (2D) light intensity patterns at a particular transverse plane using modal bases such as Laguerre-Gauss, Hermite-Gauss and Ince-Gauss modes \cite{PhysRevA.45.8185,Siegman:73,Bandres:04,Pinnell:20}, or by deploying iterative algorithms that solve inverse problems of light propagation, notably the Gerchberg–Saxton algorithm \cite{Gerchberg1972}. Although these methods provide control over the transverse field distribution at a given 2D plane, they lack control over the longitudinal beam profile. Tailoring the spatial and temporal properties of light over a 3D volume is a sought-after goal in many applications, including microscopy, spectroscopy, light-matter interactions, and optical sensing
\cite{Forbes2021,Piccardo_2022}.

A variety of wavefront shaping methods have been developed to control light's intensity in 3D. The common approach is to discretize a target 3D light distribution into a set of independent primitives. The incident waveform required to generate this target distribution is then synthesized by superposing the diffraction patterns from all the primitives in a given transverse plane. This is the principle employed in computer-generated holograms (CGH) which are often classified based on the type of their primitive \cite{Jae_recent_progress,Pi2022}. Point-cloud techniques, for instance, adopt a collection of source points, each emitting a spherical wave towards the transverse plane (CGH screen). The huge number of primitives involved in this method, however, mandates the use of look-up tables and optimization algorithms to reduce the memory consumption \cite{Tsang:18}. Although polygon mesh techniques involve a relatively smaller number of primitives, they require an additional process of shading and texture mapping to maintain the quality of the reconstructed holograms, which inherently increases their computation time \cite{Zhang:22}. Multi-plane techniques mitigate this by adopting a set of parallel planes, uniformly spaced along the propagation direction. In this case, Fresnel diffraction or angular spectrum algorithms are implemented to compute the diffraction pattern from each plane \cite{Bayraktar:10,Okada:13,Zhao:15}. These techniques have been widely adopted for creating 3D holograms as they demand a significantly smaller number of primitives compared to point-cloud and polygon mesh techniques \cite{Chen:15,Zhang:17_occlusion_effect}. However, since the planes are computed independently from each other, cross-talk between their diffraction patterns often degrades the fidelity and reconstruction quality. As a consequence, inter plane cross-talk must be reduced using optimization algorithms \cite{Makey2019,PANG2021,Velez-Zea:22,Wang:23_2,Wang:23_cross_taslk_free} which, despite their recent advances, still involve a trade-off between their computational time and reconstruction quality.

Wavefront shaping along the optical path can alternatively be achieved via a superposition of co-propagating Bessel modes with different longitudinal wavenumbers. By harnessing the spatial beating between these modes, one can modulate not only the intensity profile\cite{Michel2004,Vieira:12,Cizmar:09}, but also other degrees-of-freedom of light along the propagation direction including its total angular momentum \cite{Dorrah2016,Dorrah2021,2023D} and its polarization \cite{Dorrah2021_polariz}. By assembling many of these (nominally) non-diffracting light threads (uniformly spaced from each other) over a horizontal Cartesian plane, oriented parallel to the propagation direction, one can construct 2D light sheet whose intensity profile can be controlled at-will over the horizontal plane. By stacking several of those sheets over a volume, 3D structured light waves can also be synthesized \cite{AndreAmbrosio:19,deSarro:21, Dorrah2023}. Despite the continuous depth and axial resolution afforded by this technique, it still suffers from undesired cross-talk between the projected light sheets. Additionally, the reconstructed light fields often lack intensity uniformity owing to the transverse discretization of the co-propagating light threads. Convolutional neural networks have proven effective in reducing the cross-talk and also in enhancing the intensity uniformity \cite{Asoudegi:24}. However, this solution can be computationally demanding as it grows quadratically with the number of light sheets.

Multi-plane and light sheet holography both rely on optimization algorithms to minimize cross-talk, adding computational cost. The cross-talk in both methods exists as a result of employing waveforms (or modes) which do not represent an orthogonal basis and are not optimized to the entire 3D domain where the target object is projected. In this work, we propose and experimentally demonstrate a new wavefront shaping method that synthesizes a target 3D light distribution from the optimum basis set of orthogonal communication modes computed by means of the singular value decomposition (SVD) modal optics \cite{Miller:98}. First we define the 3D domain of the structured light field as a receiving space. Separated from this receiving space by a finite longitudinal distance, we also define a transverse plane as a source space in which the required incident waveform is encoded to create the target structured light field. Then by computing the SVD of a coupling operator connecting these two spaces, we establish communication modes, each one comprising a pair of eigenfunctions, one in the source space and another one in the receiving space. These communication modes constitute the optimum orthogonal channels connecting these spaces for the following reasons: both the source and receiving eigenfunctions form a complete orthogonal basis for their respective spaces; and each receiving eigenfunction corresponds to the largest possible magnitude of wave function its associated source eigenfunction can create in the receiving space. Therefore, the structured light waves constructed by interfering these modes represent the optimal construction of any target field distribution. In view of that, our wavefront shaping method can also serve as a framework to tell us if a given field distribution can be constructed or not. If our approach fails to create a desired field distribution, it indicates that this distribution cannot be achieved with any other wavefront shaping technique with the same optical system.

Obtaining the optimum orthogonal communication channels (modes) between two spaces using SVD is a more general concept which was introduced in the optics literature in 1998 \cite{Miller:98} and has led to many studies in wireless communications, optical fibers and optical systems \cite{Piestun:00,Miller:19}. Notably, SVD modal optics has recently enabled integrated photonic processors to determine the most efficient waves to send information through arbitrary and scattering optical media \cite{SeyedinNavadeh2024} as theoretically predicted in Ref. \citeonline{Miller:2013}. Nevertheless, while communication modes have previously been analyzed for distinct geometric configurations of source and receiving spaces including spaces comprising two rectangular volumes \cite{Miller:00}, two planar transverse apertures \cite{MARTINSSON2008,MARTINSSON2010} and even between an annular aperture and an axial line \cite{Burvall:04}, SVD modal optics has not been applied as a wavefront shaping method to create structured light waves with long depth of field in arbitrary 3D volumes such as a set of horizontal planes, oriented parallel to the optical axis.

{\section{Wavefront shaping with communication modes}}

\begin{figure}[htbp]
	\centering
	\includegraphics[width=0.95\textwidth]{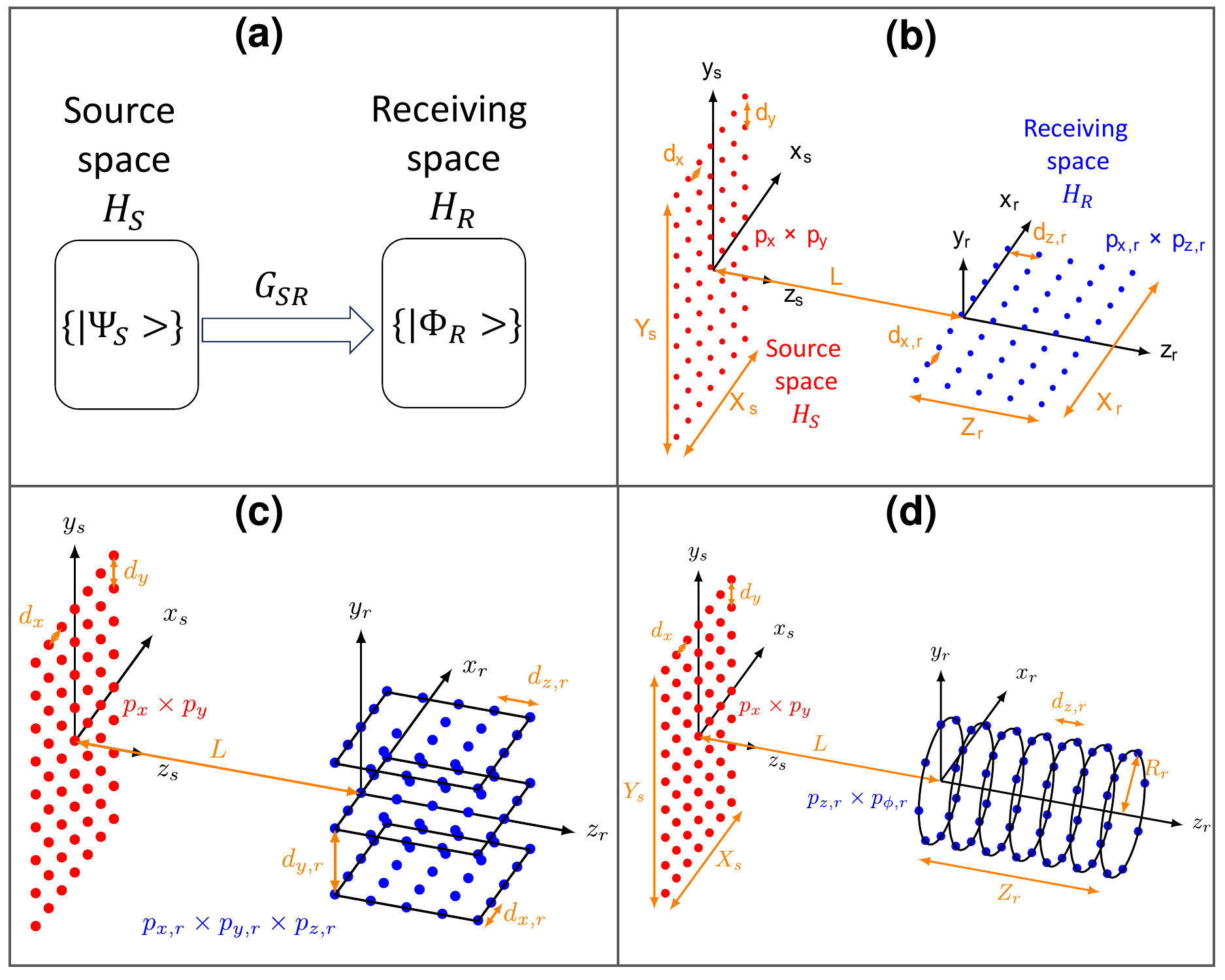}
	\caption{ \small \textbf{General concept of communication modes and examples of source and receiving spaces}. \textbf{(a)} Communication modes are established through a coupling operator $G_{SR}$ between a source and a receiving space. These spaces are mathematically viewed as Hilbert spaces ($H_S$ and $H_R$), each one containing a set of eigenfunctions ($\{\ket{\Psi_{S}}\}$ and $\{\ket{\Phi_{R}}\}$). In free space, $G_{SR}$ is expressed by a Green's function. Assuming a collection of $N_s$ source points and $N_r$ receiving points to describe these spaces, we can express $G_{SR}$ as a $N_R \times N_S$ matrix. Examples of source and receiving space configurations in which the source points are distributed as an array of $p_x \times p_y$ points in a transverse plane and the receiving points as an array of: \textbf{(b)} $p_{x,r} \times p_{z,r}$ points in a horizontal plane, \textbf{(c)} $p_{x,r} \times p_{y,r} \times p_{z,r}$ points within a set of uniformly spaced horizontal planes, and \textbf{(d)} $p_{z,r} \times p_{\phi,r}$ points disposed within a set of uniformly spaced rings placed along the optical axis. In all these configurations, the spaces are separated by a distance $L$, an on-axis distance between their coordinate systems ($x_s$,$y_s$,$z_s$) and ($x_r$,$y_r$,$z_r$).}
\label{fig_svd_concept}
\end{figure}

In SVD modal optics, the source and receiving spaces are mathematically viewed as Hilbert spaces ($H_S$ and $H_R$) that contain the possible source and receiving eigenfunctions, $\ket{\Psi_{S}}$ and $\ket{\Phi_{R}}$, as illustrated in Fig. \ref{fig_svd_concept}(a). The connection between these spaces is established through a coupling operator $G_{SR}$, which for free-space scalar waves can be described by a Green's function \cite{Miller:98}:   
\begin{equation} \label{eq_GRS}
    G_{SR,\lambda}(\textbf{r}_R,\textbf{r}_S) = - \frac{1}{4 \pi} \frac{\text{exp}(\text{i} k |\textbf{r}_R-\textbf{r}_S|)}{|\textbf{r}_R-\textbf{r}_S|} \text{,}
\end{equation}
which maps a position $\textbf{r}_S$ at the source space to a position $\textbf{r}_R$ at the receiving space for a given operating wavelength $\lambda$. Such a scalar Green's function will usually be sufficient for describing an electromagnetic wave of a single polarization. For cases of tight focusing or near field behavior, or to use this approach for full vector fields, a similar approach can be taken using the full dyadic Green's function \cite{Miller:19}. In Eq. (\ref{eq_GRS}), $k = 2\pi /\lambda$ is the wave number and a time harmonic dependence $\exp(-\text{i}\omega_0 t)$ is assumed, with $\omega_0 = k c$ being the operating angular frequency and $c$ the light speed in free space. Following Ref. \citeonline{Miller:19}, we presume that the source space consists of $N_S$ source points located at positions $\textbf{r}_{S,j}$ ($j=1,...,N_S$) while the receiving space contains $N_R$ receiving points at positions $\textbf{r}_{R,i}$ ($i=1,...,N_R$), allowing us to describe $G_{SR}$ as a $N_R \times N_S$ matrix: 
\begin{equation} \label{eq_GRS_matrix}
    g_{ij} = - \frac{1}{4 \pi} \frac{\text{exp}(\text{i} k |\textbf{r}_{R,i}-\textbf{r}_{S,j}|)}{|\textbf{r}_{R,i}-\textbf{r}_{S,j}|} \text{.}
\end{equation}

The eigenfunctions $\ket{\Psi_{S}}$ and $\ket{\Phi_{R}}$ are found by solving the SVD of the coupling operator matrix $G_{SR}$ of Eq. (\ref{eq_GRS_matrix}) which is equivalent to solving two eigenproblems, one associated with the operator $G^{\dag}_{SR} G_{SR}$ and another one associated with $G_{SR} G^{\dag}_{SR}$, leading to a one-to-one (injective) relation between these eigenfunctions and allowing us to establish the concept of a communication mode: a pair of eigenfunctions, one in the source space $\ket{\Psi_{S,j}}$ that couples to another one in the receiving space $\ket{\Phi_{R,j}}$, with the coupling strength of this connection given by the squared absolute value of the singular values $s_j$ of $G_{SR}$, i.e., $|s_j|^2$ (see Section 1 of Supplement 1). Each of these eigenfunctions is mathematically a column vector, whose elements are the amplitudes at each different point in the appropriate space. Notice that $G^{\dag}_{SR} G_{SR}$, described by a $N_S \times N_S$ matrix, is an operator within the source space, mapping a vector in $H_S$ back to another vector in $H_S$. Similarly, $G_{SR} G^{\dag}_{SR}$ is an operator within the receiving space, mapping from $H_R$ back into $H_R$, being described by a $N_R \times N_R$ matrix. Since each of these two operators is a positive Hermitian operator, its eigenfunctions (or eigenvectors) are orthogonal and form a complete set for its Hilbert space, while its eigenvalues, given by the coupling strengths $|s_j|^2$, are positive real numbers. Moreover, the sets $\{\ket{\Psi_{S,j}}\}$ and $\{\ket{\Phi_{R,j}}\}$ constitute the optimum possible orthogonal channels connecting the two spaces in terms of magnitude of the inner product. In other words, each receiving eigenfunction $\ket{\Phi_{R,j}}$ corresponds to the largest possible magnitude of wave function its associated source eigenfunction $\ket{\Psi_{S,j}}$ can create at the receiving space. For further details about this maximization property, see Section 1 of Supplement 1. Additionally note that, if the points are sufficiently dense in both spaces, this approach correctly converges towards the corresponding continuous functions and spaces, with convergence guaranteed by the Hilbert-Schmidt nature of Green's function operators for waves \cite{Miller:19}.



Given a target profile $\ket{\Phi_T}$, i.e., a vector of amplitudes at the receiving points, the required source function $\ket{\Psi_T}$ (vector of amplitudes at the source points) is determined by \cite{Miller:19} (see Section 2 of Supplement 1): 
\begin{equation} \label{eq_req_source}
    \ket{\Psi_T} = \sum_{j} \frac{1}{s_j} \braket{\Phi_{R,j}|\Phi_T} \ket{\Psi_{S,j}}\text{,}
\end{equation}
in which $\braket{\Phi_{R,j}|\Phi_T}$ is the projection of the target profile onto the set of receiving eigenfunctions. Because $\ket{\Psi_T}$ is inversely proportional to the singular values $s_j$, in practice the summation in Eq. (\ref{eq_req_source}) is truncated to the first $M$ well coupled communication modes. Since the sum of these strengths is finite, bounded by a sum rule $S$ (see Section 1 of Supplement 1), we choose a number $M$ of modes that nearly provide $S$, avoiding the incorporation of modes ($j > M$) with negligible coupling strengths. Equivalently, we can also give up adding any more terms when the resulting additional waves are so weakly coupled that we would need very large source amplitudes to generate them. The resulting wave created by the source function $\ket{\Psi_T}$ at a position $\textbf{r}$ away from the source space is given by the sum of all the spherical waves emitted by the $N_S$ source points weighted by $\ket{\Psi_T}$ \cite{Miller:19}:
\begin{equation} \label{eq_result_wave}
    \phi(\textbf{r}) = - \frac{1}{4 \pi} \sum_{q=1}^{N_S} \frac{\text{exp}(i k |\textbf{r}-\textbf{r}_{S,q}|)}{|\textbf{r}-\textbf{r}_{S,q}|} h_q \text{,}
\end{equation}
where $h_q$ is the $q$-th component of $\ket{\Psi_T}$, such that $\ket{\Psi_T} = [h_1 \hspace{0.1cm} h_2 \hspace{0.1cm} ... \hspace{0.1cm} h_{N_s}]^{T}$.



Assuming Cartesian coordinate systems ($x_s$,$y_s$,$z_s$) and ($x_r$,$y_r$,$z_r$) to describe the source and receiving spaces, we separate the origins of these two coordinate systems by a longitudinal distance $L$ and we distribute the source and receiving points in such a way that there is no overlapping between these spaces. Several examples of source and receiving point distributions are shown in Figs \ref{fig_svd_concept}(b-d). We arrange the source points as an array of $p_x \times p_y$ points in a transverse plane and space them from each other by distances $d_x$ and $d_y$ along the $x_s$ and $y_s$ directions, resulting in a source plane with dimensions $X_s = (p_x-1)d_x$ and $Y_s = (p_y-1)d_y$. In this plane, we compute the required source function $\ket{\Psi_T}$ according to Eq. (\ref{eq_req_source}). The receiving points, on the other hand, are distributed in a 2D or 3D domain where we define their amplitude values according to a target light intensity profile $|\Phi_T|^2$. In Fig. \ref{fig_svd_concept}(b), the receiving points are distributed in the horizontal plane $y_r = 0$ as an array of $p_{x,r} \times p_{z,r}$ points. They are spaced from each other by distances $d_{x,r}$ and $d_{z,r}$ along the $x_r$ and $z_r$ directions, leading to a horizontal plane with dimensions $X_r = (p_{x,r}-1)d_{x,r}$ and $Z_r = (p_{z,r}-1)d_{z,r}$. A 3D distribution of receiving points is shown in Fig. \ref{fig_svd_concept}(c) in which a set of $p_{y,r}$ horizontal receiving planes are uniformly spaced from each other by a distance $d_{y,r}$. Non-Cartesian distributions are also possible in our wavefront shaping method as depicted in Fig. \ref{fig_svd_concept}(d). In this case the receiving points are disposed in a set of $p_{z,r}$ rings placed along the optical axis, uniformly spaced from each other by a distance $d_{z,r}$, forming the lateral surface of a cylinder with radius $R_r$ and longitudinal length $Z_r$. Along each ring, we uniformly place $p_{\phi,r}$ receiving points.


The values of all the parameters we adopted for each source and receiving point distributions of Fig. \ref{fig_svd_concept} are listed in Table \ref{tab_param}. First, we fix the parameters of the receiving space and all the spacing distances. Next, we set the dimension $X_s$ of the source plane to be slightly larger than that of the receiving space $X_r$. Finally, the separation distance $L$ and the vertical dimension $Y_s$ of the source plane are chosen to satisfy a criterion on the maximum allowed value for the source spacing distances $d_x$ and $d_y$ (see Section 3 of Supplement 1). This criterion guarantees that the resulting wave created by the $N_S$ source points over the entire receiving space is essentially the same as if we had a continuous source. In our examples, we set $L = X_s$ and then $Y_s$ is determined by this criterion. Notice that the total number of horizontal planes we can allocate in the receiving space of Fig. \ref{fig_svd_concept}(b) is constrained by our choice on the spacing distance $d_{y,r}$. The value of $d_{y,r} = 15\lambda$ listed in Table \ref{tab_param} for this distribution is the minimum distance that we can set such that the reconstructed intensity profiles in all the horizontal planes using our wavefront shaping approach do not present substantial cross-talk. This fact, which is analyzed in detail in the Results section, indicates that structuring light waves in these horizontal planes with a tighter spacing distance between them ($d_{y,r} < 15\lambda$) is inevitably constrained by diffraction limits and thus significant cross-talk is always present regardless of the wavefront shaping technique and optimization techniques we adopt.

\begin{table}
\caption{Number of source and receiving points, their spacing distances and the longitudinal separation between the source and receiving spaces for each example of source and receiving space configuration of Fig. \ref{fig_svd_concept}. For the example of Fig. 1(d), we present the radius $R_r$ of each ring that compose the receiving surface. The separation distance of the receiving points along each ring is $d_{\phi,r} = (2\pi/p_{\phi,r})R_r =  1.55 \lambda$, which corresponds to an angular separation of $3.56^\circ$.}
\centering
\label{tab_param}
\begin{tabular}{lllllll}
\cline{1-6}
\multicolumn{1}{|c|}{\multirow{2}{*}{Configuration}} & \multicolumn{2}{c|}{Source Plane}                         & \multicolumn{1}{c|}{\multirow{2}{*}{$L$}} & \multicolumn{2}{c|}{Receiving space}                                                                                     &  \\ \cline{2-3} \cline{5-6}
\multicolumn{1}{|c|}{}                              & \multicolumn{1}{c|}{$p_x \times p_y$} & \multicolumn{1}{c|}{$d_x$,$d_y$} & \multicolumn{1}{c|}{}                   & \multicolumn{1}{c|}{\begin{tabular}[c]{@{}c@{}}array of \\ points\end{tabular}} & \multicolumn{1}{c|}{spacing distances} &  \\ \cline{1-6}
\multicolumn{1}{|c|}{Fig.1(b)}                             & \multicolumn{1}{c|}{$111 \times 222$}     & \multicolumn{1}{c|}{$\lambda$}    & \multicolumn{1}{c|}{110$\lambda$}                 & \multicolumn{1}{c|}{$p_{x,r} \times p_{z,r} = 101 \times 101$}                                                        & \multicolumn{1}{c|}{$d_{x,r} = d_{y,r} = d_{z,r} = \lambda$}                &  \\ \cline{1-6}
\multicolumn{1}{|c|}{Fig.1(c)}                             & \multicolumn{1}{c|}{$111 \times 301$}     & \multicolumn{1}{c|}{$0.5\lambda$}    & \multicolumn{1}{c|}{50$\lambda$}                & \multicolumn{1}{c|}{\begin{tabular}[c]{@{}c@{}}$p_{x,r} \times p_{z,r} = 51 \times 51$ \\ $p_{y,r} = 10$\end{tabular}}                                                        & \multicolumn{1}{c|}{\begin{tabular}[c]{@{}c@{}}$d_{x,r} = d_{z,r} = \lambda$ \\ $d_{y,r} = 15\lambda$\end{tabular}}                &  \\ \cline{1-6}
\multicolumn{1}{|c|}{Fig.1(d)}                             & \multicolumn{1}{c|}{$201 \times 201$}     & \multicolumn{1}{c|}{$0.5\lambda$}    & \multicolumn{1}{c|}{50$\lambda$}                & \multicolumn{1}{c|}{\begin{tabular}[c]{@{}c@{}}$p_{z,r} = 101$ \\ $p_{\phi,r} = 101$\end{tabular}}                                                        & \multicolumn{1}{c|}{\begin{tabular}[c]{@{}c@{}}$d_{z,r} = 0.5\lambda$ \\ $R_{r} = 25 \lambda$\end{tabular}}                &  \\ \cline{1-6}
\end{tabular}
\end{table}

\begin{figure}[htbp]
	\centering	\includegraphics[width=1\textwidth,height=0.8\textwidth]{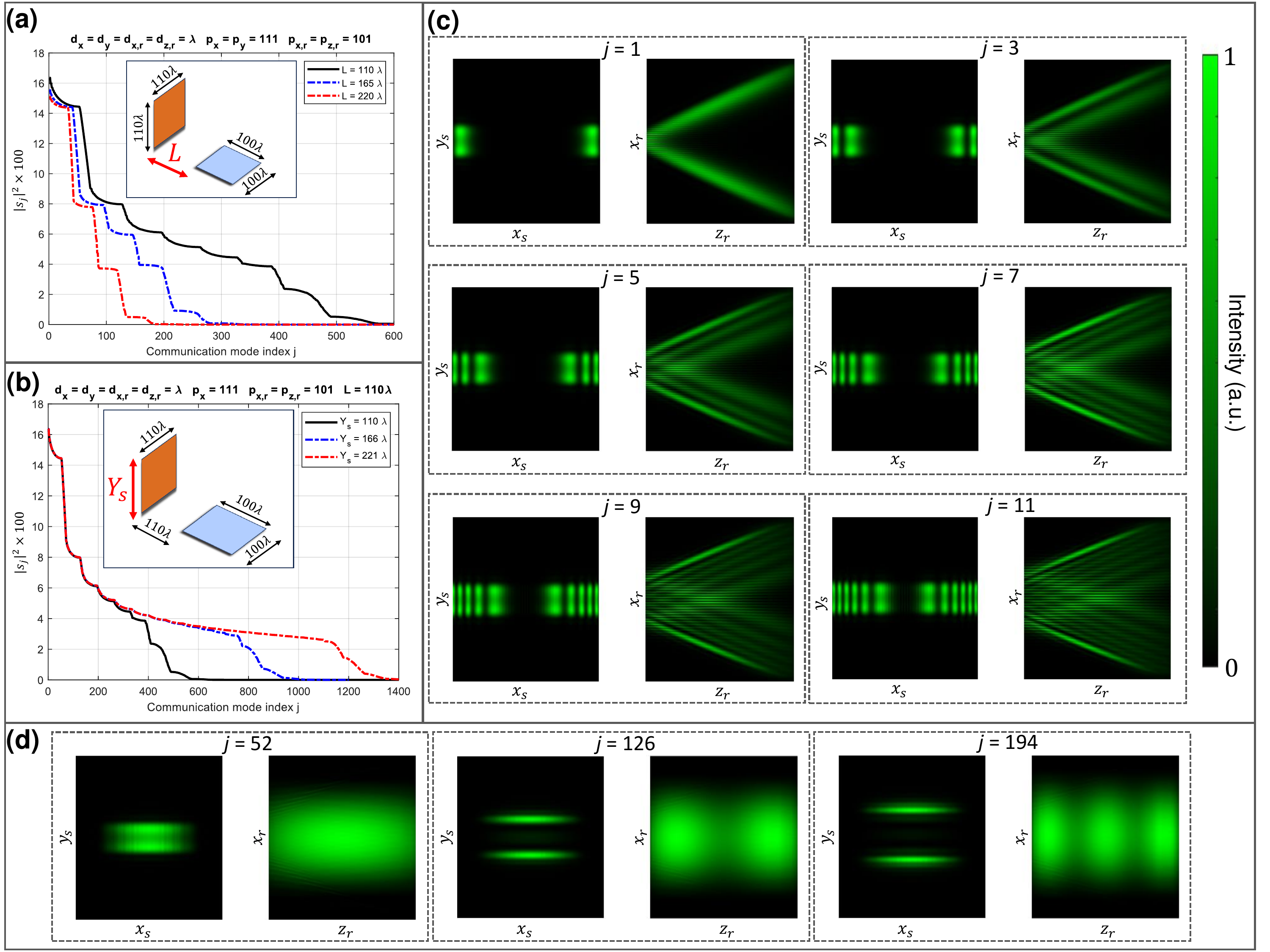}
	\caption{\small \textbf{Communication modes and their coupling strengths associated with a transverse source plane and a horizontal receiving plane}. Coupling strengths in order of decreasing size of their magnitude computed at $\lambda$ = 532 nm for \textbf{(a)} different values of the longitudinal separation distance $L$ between the spaces and \textbf{(b)} for different values of the source plane $y$-dimension $Y_s = (p_y-1) d_y$. The other parameters, remained fixed, are shown at the title of each sub-figure and follows Table 1. The coupling strength curves are closely characterized by a series of steps after which the singular values decrease in a rapid fall-off fashion. The width of each step and the number of steps corresponds respectively to the number of effective transverse and longitudinal modes. Normalized squared amplitude of \textbf{(c)} the first six odd-numbered communication modes  and \textbf{(d)} the last mode of each of the first three steps of the red dashed line of (b). In each sub-figure in (c) and (d), the source eigenfunction is shown on the left and its associated receiving eigenfunction on the right.}
\label{Fig_2D_cart_eigenvalues}
\end{figure}

The coupling strengths $|s_j|^2$ in order of decreasing size of their magnitude associated with the distribution of Fig. \ref{fig_svd_concept}(a), computed at $\lambda$ = 532 nm, are shown in Fig. \ref{Fig_2D_cart_eigenvalues}(a) for different values of the separation distance $L$ and in Fig. \ref{Fig_2D_cart_eigenvalues}(b) for different values of the source plane $y$-dimension $Y_s$. In Fig. \ref{Fig_2D_cart_eigenvalues}(c) we show the normalized intensity profile of the first six odd-numbered communication modes associated with the red dashed curve of Fig. \ref{Fig_2D_cart_eigenvalues}(b) at the source plane (on the left of each sub-figure) and at the receiving horizontal plane (on the right). As expected for this distribution, the intensity profiles of the source eigenfunctions are symmetric with respect to the $y_s = 0$ axis. Additionally, the receiving eigenfunctions do not correspond to any of the standard beams (e.g., Bessel, Airy beams). Finally, notice that the coupling strength curves are closely characterized by a series of steps. The width of each step, which is increased by reducing the distance $L$, corresponds to the number of effective transverse modes along the receiving plane $x_r$ direction. On the other hand, the number of steps, which is increased by either reducing the distance $L$ or increasing the source plane $Y_s$ dimension, corresponds to the number of effective longitudinal modes along the $z_r$ direction. This is illustrated in Fig. \ref{Fig_2D_cart_eigenvalues}(d) which shows the normalized intensity profile of the last mode of each of the first three steps of the red dashed curve of Fig. \ref{Fig_2D_cart_eigenvalues}(b). After the series of steps, the singular values decrease in a rapid fall-off fashion. This kind of fall-off, which limits the number of usable modes, is a universal behavior, as detailed in Ref. \citeonline{Miller2024}. A complete analysis of the intensity profiles of the communication modes and their coupling strengths for all the distribution examples of Fig. \ref{fig_svd_concept} is provided in Section 4 of Supplement 1. Additionally, the intensity profiles of the first well-coupled communication modes associated with these distributions are shown in Supplementary Videos 1, 2 and 3.

Next we apply Eqs. (\ref{eq_req_source}) and (\ref{eq_result_wave}) to structure waves in the receiving space. We first consider the configuration of the receiving and source space of Fig. \ref{fig_svd_concept}(b) parameterized according to Table \ref{tab_param}. The dimensions of these spaces are depicted in Fig. \ref{Fig_exp_gen}(a). We also consider the binary 2D image shown in Fig. \ref{Fig_exp_gen}(b) as our target receiving intensity profile $|\Phi_T|^2$. To project the entire content of the target amplitude profile onto the range of well-coupled modes of the receiving space set, we first need to modulate the target amplitude profile with a phase front of the form $\exp(\text{i} Q z_r)$, in which $Q$ is a positive real constant with $Q \leq k$. We do this because we understand that any wave we generate from our source plane is physically going to have predominantly some underlying phase variation along the $z$ direction, and so we can only reasonably generate patterns that have a similar underlying phase variation. A plane wave phase form is the simplest one that allows us to generate the intensity patterns of interest using only strongly coupled communication modes. This is demonstrated in Fig. \ref{Fig_exp_gen}(c) in which the blue dots represent the amplitude of the inner product coefficients between the receiving eigenfunctions $\{\ket{\Phi_{R,j}} \}$ and the modulated target profile $\Phi_T \exp(\text{i} Q z_r)$, with $Q = 0.95 k$. For reference, we also show the coupling strengths $|s_j|^2$ (in arbitrary units) in the red dashed line. The squared amplitude of the required source function $|\Psi_T|^2$ and the resulting wave intensity $|\phi|^2$, calculated using the first 1200 modes ($j \leq 1200$), are shown in Fig. \ref{Fig_exp_gen}(d). Incorporation of the phase front $\exp(\text{i} Q z_r)$ results in a source function profile $|\Psi_T|^2$ with an increased extent $\Delta Y$ along the source plane $y_s$ dimension, capable of rendering waves with high longitudinal spatial frequencies within the receiving space, giving rise to a resulting wave that fully resolves the target amplitude profile $|\Phi_T|$. In Section 5 of Supplement 1 we relate the extent $\Delta Y$ of the source profile with the characteristic minimum length of wave field (minimum spot) $\Delta z$ that we can create along the longitudinal direction at a certain position $z_0$ from our finite source transverse aperture. Additionally, the incorporation of the plane wave phase front is analyzed for different values of $Q$ in Section 6 of Supplement 1.

\subsection{Experimental Generation}

We optically reconstructed light waves at operating wavelength $\lambda$ = 532 nm using a phase-only reflective SLM (SANTEC 200) with pixel pitch $\delta$ = 8 $\mu$m and 1,920 x 1,200 resolution. Although the SLM is a highly tunable platform that allows us to easily test our method for different target intensity profiles by simply modifying its encoded phase mask, its large pixel pitch $\delta \approx 15 \lambda$ demands a source transverse plane with dimensions of thousands of wavelengths. For an effective SLM display area of 600 x 600 pixels, the array of source and receiving points (and the separation distance $L$) of the distribution of Fig. \ref{fig_svd_concept} parametrized as listed in Table \ref{tab_param} need to be scaled up by a factor of $\alpha = 40.5$. Since the SVD computation (communication modes and their coupling strengths) of this scaled distribution is not feasible (see Section 7 of Supplement 1 for details), we instead opt to encode in the SLM the wave solutions computed from the distributions listed in Table \ref{tab_param}. A side effect of this approach is that the 1:1 aspect ratio of our original (simulated) light wave within the receiving space is not preserved in our optically reconstructed light waves, as the wave solutions are significantly scaled to match the SLM effective display area, resulting in waves that are substantially stretched along the propagation direction $z_r$. This, of course, is not a fundamental limitation of our wavefront shaping method but rather a technical one because of the use of SLMs. Metasurfaces with their subwavelength pixel pitch would allow us to encode directly the required source function without any magnification, thus preserving the original 1:1 aspect ratio. However, a drawback of a metasurface is its lack of tunability. Different metasurfaces need to be fabricated to create distinct target intensity profiles. For details on the aspect ratio of our optically reconstructed light waves using the SLM and how a 1:1 aspect ratio could be achieved by means of a de-magnification 4$f$ system, see Section 8 of Supplement 1.

\begin{figure}[htbp]
	\centering
	\includegraphics[width=0.8\textwidth,height=0.85\textwidth]{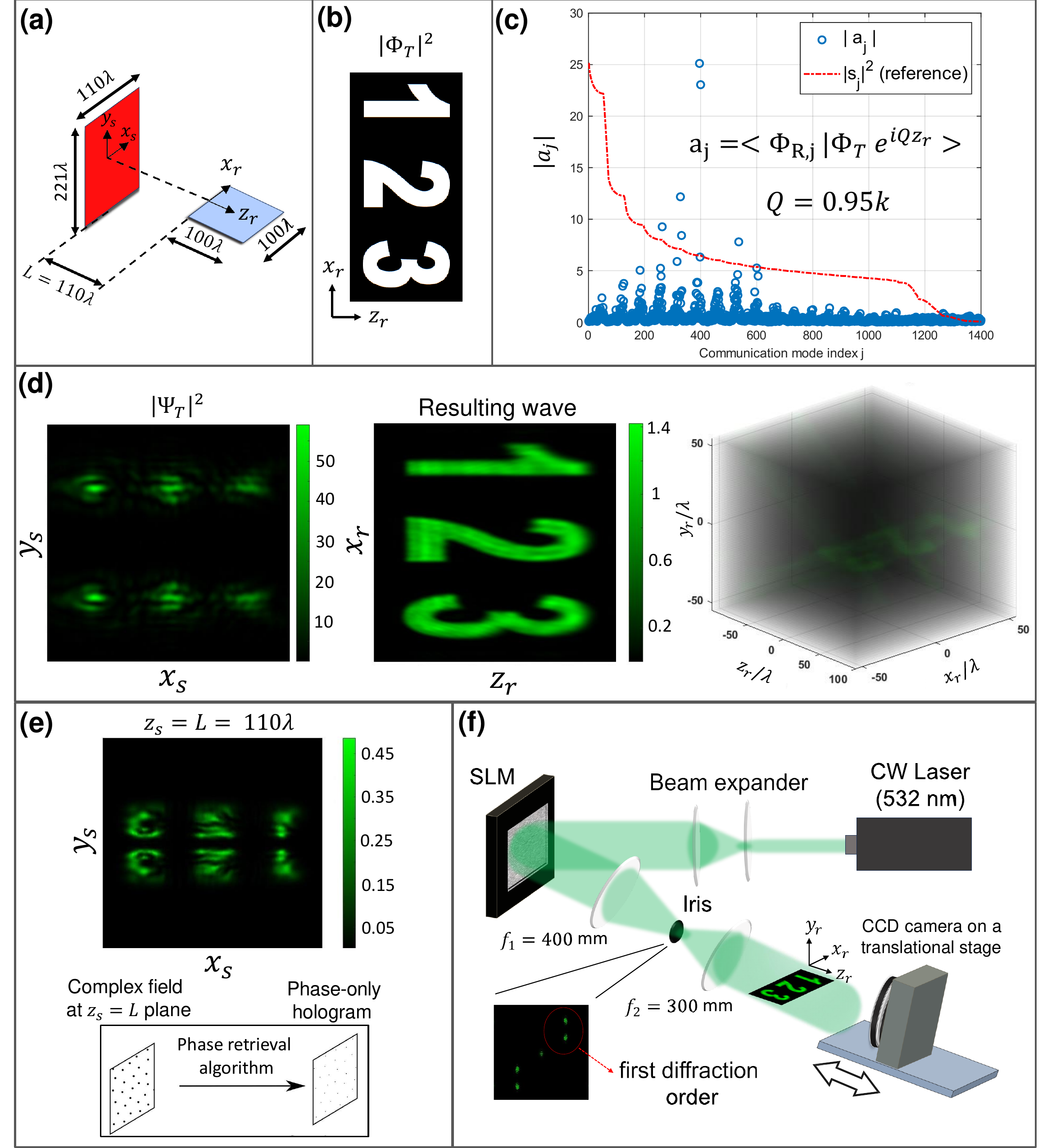}
	\caption{ \small \textbf{Optical reconstruction of structured light waves based on communication modes using a reflective phase-only SLM}. For a given source and receiving space configuration \textbf{(a)}, we compute the associated communication modes and their coupling strengths. \textbf{(b)} We define a target light intensity distribution \textbf{(c)} and project it onto the set of receiving eigenfunctions. \textbf{(d)} We compute the required source function $\Psi_T$ at the source plane and its associated resulting wave $\phi$ within the receiving space. \textbf{(e)} Encoding process: we evaluate the resulting wave from the required source profile at the plane $z_s = L$, scale it up to match the SLM effective display area and convert its complex field distribution into a phase-only CGH by means of a phase retrieval algorithm. \textbf{(f)} Optical setup: after the SLM, we employ a standard 4$f$ lens system to recover the complex field $\phi(x_s,y_s,L)$ at front focal plane of the second lens. At the Fourier plane (mutual focal point plane of the two lenses), an iris is used to filter the desired spectrum (encoded on the first diffraction) and to block the unmodulated SLM zeroth diffraction order. A CCD camera on a translational stage records transverse intensity distributions of the resulting wave at different longitudinal positions.}
\label{Fig_exp_gen}
\end{figure}

To encode the wave solutions computed from the distributions listed in Table \ref{tab_param} in the SLM, first, as illustrated in Fig. \ref{Fig_exp_gen}(e), we compute the resulting wave at the plane $z_s = L$, i.e., $\phi(x_s,y_s,L)$ from Eq. (\ref{eq_result_wave}) and we scale it up to match the SLM effective display area (600 x 600 pixels). We then convert its complex field into a phase-only mask using a phase retrieval algorithm (see Section 9 of Supplement 1). Finally, we add a blazed grating profile to the phase-only mask to operate off-axis, separating the hologram's spectrum from the SLM zeroth diffraction order in $k$ space. Our experimental setup is shown in Fig. \ref{Fig_exp_gen}(f). The laser beam is first expanded and collimated to illuminate the entire SLM effective screen as a uniform plane-like wave. After the SLM, we place a standard 4$f$ lens system whose purpose is to recover the complex field $\phi(x_s,y_s,L)$ at the front focal plane of the second lens. The first lens performs a Fourier transform operation, projecting the phase mask spectrum (which contains both amplitude and phase information) at the Fourier plane, i.e., at the mutual focal point plane of the two lenses. In this plane, we place an iris to spatially filter the desired spectrum, encoded on the first diffraction order of the incoming beam, while blocking the undesired zeroth diffraction order in $k$-space. The second lens performs an inverse Fourier operation, transforming the filtered spectrum into real space, recovering the complex field $\phi(x_s,y_s,L)$ in the front focal plane of this lens. Using a charge-coupled device ($\mu$Eye UI224SE-M, 1,280 x 1,024 resolution and 4.65 $\mu$m pixel size), mounted on a translational stage (Thorlabs LTS150), we record transverse intensity distributions of the resulting wave at different $z_r$ positions. Stacking these transverse distributions, we obtain the optical reconstruction of the resulting wave intensity.

\section{Results}

\begin{figure}[htbp]
	\centering
	\includegraphics[width=0.65\textwidth,height=0.9\textwidth]{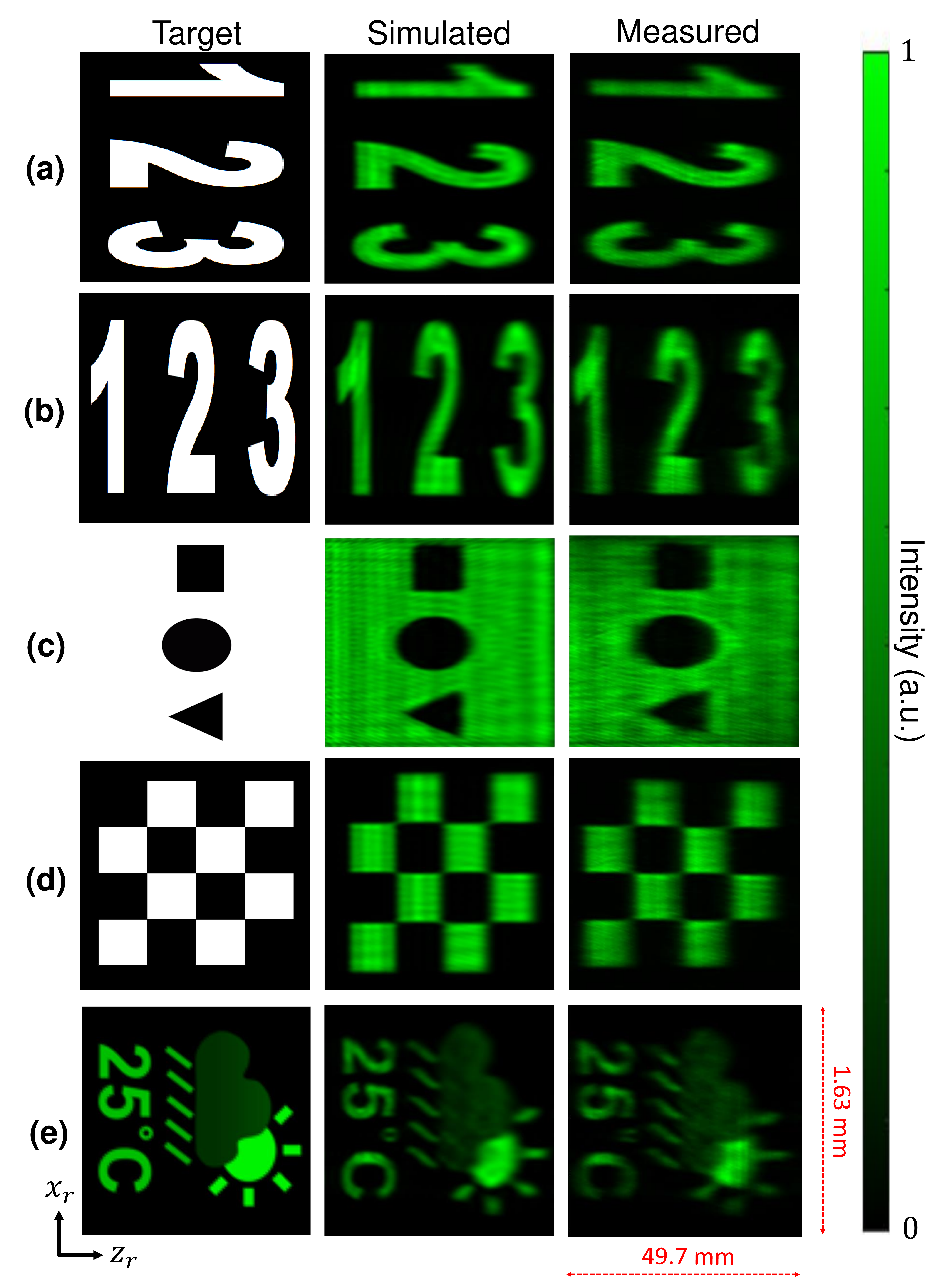}
	\caption{ \small \textbf{Examples of 2D structured light waves projected in a horizontal receiving plane}. The structured light waves are computed from the first 1200 well coupled communication modes associated with the source and receiving space configuration of Fig. \ref{fig_svd_concept}(b) with parameters listed on Table \ref{tab_param}. From left to right: target receiving amplitude, simulated resulting wave intensity at the receiving horizontal plane, and optical reconstruction of the resulting wave using a phase-only SLM. \textbf{(a-b)} Bright digits on a dark background with distinct orientations. \textbf{(c)} Dark geometric shapes on a bright background. \textbf{(d)} A checkerboard pattern. \textbf{(e)} A grayscale image. Since the wave solutions are scaled up to match the SLM effective display area, the 1:1 aspect ratio of the simulated waves is not preserved in the measured results. Their measured dimensions are $Z_r \times X_r$ = 49.7 mm $\times$ 1.63 mm.}
\label{fig_svd_2D_holograms}
\end{figure}

For all the examples shown in this section, the projection of the target profile $\Phi_T$ onto the set of receiving eigenfunctions is done by modulating $\Phi_T$ with a phase front $\exp(\text{i} Q z_r)$ with $Q = 0.95k$. First, examples of different 2D structured light waves using the first 1200 communication modes associated with the source and receiving space configuration of Fig. \ref{Fig_exp_gen}(a) are compiled in Fig. \ref{fig_svd_2D_holograms}. On the left of each subfigure, we show the target receiving amplitude $|\Phi_T|$; in the middle, the simulated resulting wave intensity $|\phi|^2$ in the receiving horizontal plane; and on the right, the optical reconstructed wave using the optical setup of Fig. \ref{Fig_exp_gen}(f). Figs. \ref{fig_svd_2D_holograms}(a-b) exhibit examples of target binary images containing bright digits on a dark background, while examples involving dark geometric shapes on a bright background and a checkerboard pattern are shown in Figs \ref{fig_svd_2D_holograms}(c-d). These four examples demonstrate that our wavefront shaping method based on optimal orthogonal communication modes is able to provide high intensity uniformity in both dark and bright regions, resulting in reconstructed patterns with high contrast. Finally, in Fig.  \ref{fig_svd_2D_holograms}(e), we provide an example involving a target grayscale image containing features with high spatial frequencies. Notice that the fine rectangular features located in the middle of the receiving plane (which represent the rain drops) present a better reconstruction than the rectangular features located at the end of the receiving plane (that represent the sun rays). This phenomenon is intrinsically due to the finite dimension $Y_s$ of our source aperture and is independent of the modes (or basis functions) used to create the structured light wave. In particular, the characteristic minimum length of wave field $\Delta z$ that our (finite) source aperture can create varies along the longitudinal position $z_r$ of the receiving space, being higher for points located further away from the source plane (see Section 5 of Supplement 1). To fully resolve these fine features in our wavefront shaping method, we would need to consider a source plane with a higher dimension $Y_s$ and with a denser array of source points for this additional increment in the dimension $Y_s$ to satisfy the criteria for the source spacing distances (Section 3 of Supplement 1). Additionally, the target profile must be modulated by a plane wave phase front $\exp(\text{i} Q z_r)$ with a lower value of $Q$ so that the required source function occupies a higher vertical position in the source plane (Section 6 of Supplement 1).

\begin{figure}[htbp]
	\centering
	\includegraphics[width=0.7\textwidth,height=0.9\textwidth]{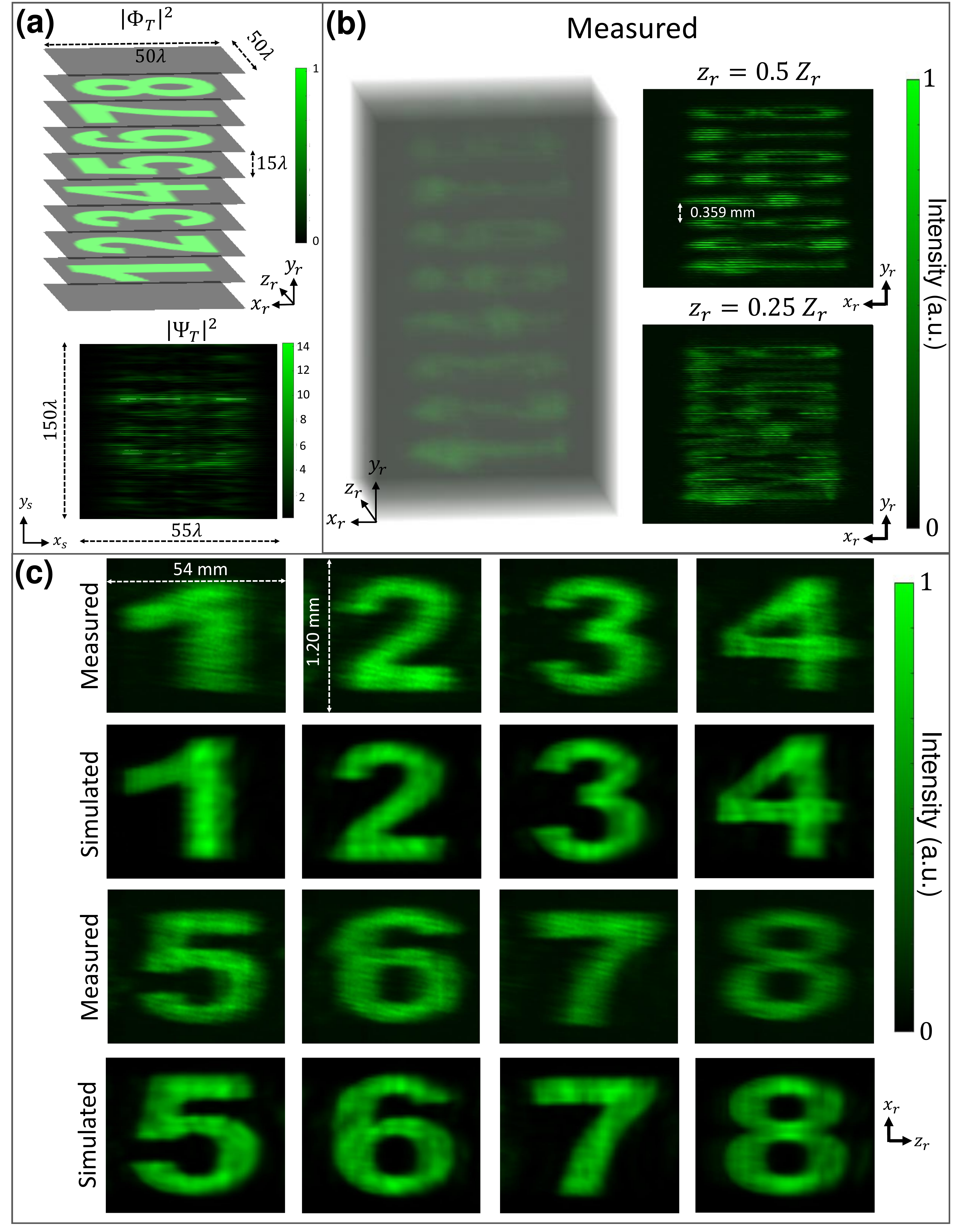}
	\caption{ \small \textbf{Example of a 3D structured  light wave projected in a set of uniformly spaced horizontal receiving planes}. Projecting eight digits within the eight inner horizontal receiving planes of the source-receiving space system of Fig. \ref{fig_svd_concept}(c) with parameters listed on Table \ref{tab_param}. The required source function and the resulting wave are computed using the first 3500 modes. \textbf{(a)} Target receiving profile $|\Phi_T|^2$ intensity and corresponding required source function $|\Psi_T|^2$ intensity. \textbf{(b)} Optical reconstruction of the resulting wave using a phase-only SLM within a volume containing the horizontal receiving planes and at transverse planes located at the mid and quarter longitudinal distances, $z_r = Z_r/2$ and $z_r = Z_r/4$ ($Z_r$: longitudinal length of the planes). The wave solution is scaled up to match the SLM effective display area and thus the 1:1 aspect ratio of the simulated wave is not preserved in the measured results. Measured spacing distance between the horizontal planes is $d_{y,r} = 0.359$ mm while their dimensions are $Z_r \times X_r$ = 54 mm $\times$ 1.20 mm. \textbf{(c)} Measured and simulated results of all eight reconstructed digits.}
\label{fig_8_digits}
\end{figure}

Next we present examples of 3D structured light waves computed using the first 3500 modes associated with the source and receiving space configuration of Fig. \ref{fig_svd_concept}(c) and parameterized as listed in Table \ref{tab_param}. First, we project eight distinct digits, from '1' to '8', assigning each digit binary amplitude profile to one of the eight inner horizontal receiving planes. In Fig. \ref{fig_8_digits}(a) we show the target receiving intensity profile $|\Phi_T|^2$ and the intensity of the corresponding required source function $|\Psi_T|^2$. The optical reconstruction of the resulting wave using a phase-only SLM is shown in Fig. \ref{fig_8_digits}(b), in which we can clearly identify and distinguish all eight digits, indicating the low level of crosstalk between their reconstructed intensity profiles, which can also be evidenced from transverse planes of the optical reconstruction as seen in the same figure at the mid ($z_r = Z_r/2$) and quarter ($z_r = Z_r/4$) longitudinal distances. In Fig. \ref{fig_8_digits}(c), we compare the measured and simulated results of the structured light wave intensity for all eight inner horizontal receiving planes. Notice that all the digits are reconstructed with high fidelity, presenting high intensity uniformity and contrast. An additional example of a structured light wave involving the projection of eight layers of an ellipsoid is provided and discussed in Section 10 of Supplement 1. All these examples illustrate how powerful our wavefront shaping method is in creating structured light waves with continuous depth of field, high fidelity, and negligible cross-talk in a set of horizontal planes uniformly spaced by a distance of only 0.359 mm. In contrast, creating light waves in this set of planes using the light sheet wavefront shaping method results in reconstructed intensity profiles with a high level of cross-talk. This is demonstrated in Section 11 of Supplement 1, in which we compare our method with the light sheet one by evaluating the level of cross-talk in all eight horizontal planes of the example of Fig. \ref{fig_8_digits}.

Our wavefront shaping method based on communication modes can be used as a framework to tell us if a given field distribution can be constructed or not. Since the communication modes represent complete orthogonal sets of optimally well-coupled modes that connect the source and receiving spaces, if we cannot construct a given desired field using communication modes above some practical minimum coupling strength, then we cannot construct that field by any other method that uses the same optical system. To illustrate this, in Section 12 of Supplement 1 we analyze the resulting wave computed from the same source and receiving distribution in Fig. \ref{fig_8_digits} but for smaller values of spacing distances between the horizontal receiving planes ($d_{y,r} = 10 \lambda$ and $d_{y,r} = 7.5 \lambda$). Notice that the reconstructed intensity profiles are affected by cross-talk. This indicates that structuring light waves in these horizontal receiving planes for a spacing distance $d_{y,r} < 15 \lambda$ is constrained by diffraction limits and cross-talk is the effect imposed by these limits in the reconstructed intensity profiles. Regardless of the wavefront shaping technique and optimization techniques we apply to structure waves in this case, cross-talk cannot be removed. Those diffraction limits are represented in our approach by the rapid fall-off in coupling strengths of the communication modes. Indeed, the communication mode picture can be viewed as a generalization of diffraction theory to arbitrary source and receiver volumes \cite{Miller:19}. Finally, an example of a structured light wave projected onto the cylindrical surface of Fig. \ref{fig_svd_concept}(d) and parameterized as listed in Table \ref{tab_param} is presented and discussed in Section 13 of Supplement 1. While for a radius of $R_r = 25 \lambda$ the resulting wave is concentrated only around the cylindrical surface, when this radius is reduced by half ($R_r = 12.5 \lambda$), diffraction limits impose a high intensity inside the cylindrical surface along the optical axis, which can be undesirable as it can affect the visualization of the structured light wave projected on the cylinder.

\section{Discussion}

We introduced and experimentally verified a new wavefront shaping method based on interfering the optimum orthogonal communication modes connecting a source plane and a receiving volume. Each communication mode encompasses a pair of eigenfunctions, one in the source space and another one in the receiving space, both computed by means of the singular value decomposition modal optics. The orthogonality and optimization properties of the source and receiving eigenfunctions over the entire source and receiving spaces provide 2D and 3D structured light waves with high fidelity, high contrast and with minimal level of cross-talk, allowing these structured light waves to be perceived from different angles and thus being potentially employed in volumetric displays and AR/VR headsets. Additionally, our method could be favorable in applications that require controlling light's intensity within a 3D domain with high precision, notably in materials processing and in optical trapping. For instance, our method can potentially be employed to manipulate many micro-particles simultaneously in a set of horizontal planes with precise control. Our method can also be suitable for real-time holography or applications that require dynamic wavefront shaping. This is aided by the low computational cost associate with our method. The coupled modes connecting the source and receiver spaces are computed once, then any target object can be synthesized through a Fourier-like matrix operation which can be parallelized using GPUs.

Although we limited our analysis to a fixed operating wavelength, our method can be extended to create spatiotemporal structured light waves by considering a full time-dependent Green's function for the coupling operator. Furthermore, light waves with spatially varying polarization states can also be created by adopting a dyadic vector potential Green's function instead of a scalar one \cite{Miller:19}. In all these cases, metasurfaces \cite{Yu2014,science_abi6860} are an ideal wavefront shaping platform, as they offer multi-wavelength control and polarization transformations at the nanoscale \cite{science_abi6860}. The sub-wavelength regime of the flat-optics platform also benefits our wavefront shaping method in creating reconstructed light waves with the same aspect ratio designed for the target 3D light distribution. Therefore, in a future work, our wavefront shaping method can be implemented with metasurface platforms to achieve control over other degrees-of-freedom of light, unlocking new applications in digital holography and structured light, as well as light-matter interaction, classical and quantum communications, and beyond. Finally, because our method is based on the interfering of communication modes that leads to the optimal construction of any target field distribution, it can also be used as a benchmark to analyze the limitations of existing wavefront shaping techniques.

\begin{backmatter}
\bmsection{Acknowledgments}
The authors acknowledge the insightful discussions with Prof. M. Brongersma from Stanford University. V.S.A acknowledges financial support from the Coordination of Superior Level Staff Improvement (CAPES), grant no. 88887.833874/2023-00, and from the National Council for Scientific and Technological Development (CNPq), grant no. 140270/2022-1. A.H.D acknowledges the support of the Optica Foundation. L.A.A. acknowledges financial support from the National Council for Scientific and Technological Development (CNPq), grant no. 309201/2021-7, and from the São Paulo Research Foundation (FAPESP), grant no. 2020/05280-5. F.C. acknowledges financial support from the Office of Naval Research (ONR) under the MURI programme, grant no. N00014-20-1-2450, and from the Air Force Office of Scientific Research (AFOSR) under grant nos FA9550-21-1-0312 and FA9550-22-1-0243. D. M. also acknowledges support from AFOSR grant FA9550-21-1-0312.

\bmsection{Competing Interests}
The authors declare no competing interests.

\bmsection{Data Availability}
All key data supporting the findings of this study are included in the main article and its supplementary information. Additional data sets and raw measurements are available from the corresponding author upon reasonable request.

\bmsection{Supplemental document}
See Supplement 1 for supporting content. 

\end{backmatter}


\bibliography{main}






\end{document}